\documentclass[12pt,epsf,amssymb]{article} \textheight =23 truecm
\def\lsim{\raise0.3ex\hbox{$<$\kern-0.75em\raise-1.1ex\hbox{$\sim$}}}
\def\gsim{\raise0.3ex\hbox{$>$\kern-0.75em\raise-1.1ex\hbox{$\sim$}}}
\def\noi{\noindent}  \def\bea{\begin{eqnarray}}
\def\eea{\end{eqnarray}} \def\beq{\begin{equation}}
\def\eeq{\end{equation}} 
\def\beeq{\begin{eqnarray}} \def\eeeq{\end{eqnarray}} \def\R{ {\rm R
\kern -.31cm I \kern .15cm}} \def\C{ {\rm C \kern -.15cm \vrule
width.5pt \kern .12cm}} \def\Z{ {\rm Z \kern -.27cm \angle \kern
.02cm}} \def\N{ {\rm N \kern -.26cm \vrule width.4pt \kern .10cm}}
\def\1{{\rm 1\mskip-4.5mu l} }

\usepackage{graphics} \usepackage{graphicx} \usepackage{epsfig}

\begin{document}

\begin{center} 

{\large \bf Note on new interesting baryon channels\par to measure the photon polarization in $b \to s \gamma$}

\par \vskip 1.2 truecm

{\bf L. Oliver}$^1$, {\bf J.-C.
Raynal}$^1$ {\bf and R. Sinha}$^2$ 

\par \vskip 1 truecm

$^1$ Laboratoire de Physique Th\'eorique\footnote{Unit\'e Mixte de
Recherche UMR 8627 - CNRS }\\    Universit\'e de Paris XI,
B\^atiment 210, 91405 Orsay Cedex, France 

\vskip 0.3 truecm

$^2$ Institute of Mathematical Sciences\\ CIT Campus, Tharamani, Chennai 600113, India 

\end{center}

\vskip 0.5 truecm

\begin{abstract} 

At LHC a large number of b-flavored baryons will be produced. In this note we propose new baryon modes to determine the photon helicity of the penguin transition $b \to s \gamma$. The decay $\Lambda_b \to \Lambda \gamma$ has the drawback that the $\Lambda$, being neutral and long-lived, will escape detection most of the time. To overcome this difficulty, transitions of the type $\Lambda_b \to \Lambda^{*} \gamma$ have been proposed, where $\Lambda^{*}$ denotes an excited state decaying strongly within the detector into the clean mode $p K^-$. The doublet $\Xi_b$, that decays weakly, has a number of good features. The charged baryon $\Xi_b^-$ will decay into the mode $\Xi^- \gamma$, where the ground state hyperon $\Xi^-$, although it will decay most of the time outside the detector, can be detected because it is charged. We consider also the decay of $\Xi_b$ into $\Xi^{*} \gamma$, where a higher mass state $\Xi^{*}$ can decay strongly within the detector. We point out that the initial transverse polarization of $\Xi_b$ has to be known in all cases. To determine this parameter through the transition $\Xi_b \to J/\Psi\ \Xi$, we distinguish between different cases, and underline that in some situations one needs {\it theoretical input} on the asymmetry parameter $\alpha_{\Xi_b}$ of the primary decay. {\it A fortiori} the same considerations apply to the case of the $\Lambda_b$.

\end{abstract}

\vskip 0.5 truecm

\noi LPT Orsay 10-17 \qquad IMSc/2010/07/11 \qquad June 2010

\par \vskip 1 truecm

\newpage \pagestyle{plain}

\section{Introduction}

The photon polarization in $b \to s \gamma$ transitions is an important observable in view of detecting possible New Physics (NP). A number of processes have been proposed in the literature as $B_d \to K^*\gamma$ \cite{AGS} or more generally $B_d \to K \pi \pi \gamma$ \cite{GP}, the latter involving quasi two-body modes and a continuum. On the other hand, the baryon decay $\Lambda_b \to \Lambda \gamma$ has also been proposed \cite{GKS}\cite{MR}\cite{HK}, but it has the important drawback that the very energetic final $\Lambda$, at LHC for example, will escape detection most of the time. To overcome this difficulty there has been the proposition \cite{LS}\cite{HKLS} of decays $\Lambda_b \to \Lambda^{*} \gamma$, where $\Lambda^{*}$ is an excited state decaying of spin $1 \over 2$ or $3 \over 2$, strongly decaying within the detector into $p K^-$, a very clean signature. \par

The present note is a very straightforward application of the formalism clearly exposed in the articles by F. Legger and T. Schietinger \cite{LS} and G. Hiller et al. \cite{HKLS}. The main formulas that we need can be found in these papers. For completeness and clarity we will recall the main needed ones.\par

In this paper, instead of excited $I = 0$ baryons considered in \cite{LS}, we focus on other {\it ground state} b-flavored baryons, namely the isospin doublet $\Xi_b^0$ and $\Xi_b^-$, of spin-parity ${1 \over 2}^+$, made out of $bsu$ and $bsd$ quarks respectively, that offer, as we will see below, a number of advantages. To our knowledge, these modes have not been considered in the literature as far as the photon helicity measurement in the $b \to s \gamma$ transition is concerned. \par

The interest of the isospin doublet $\Xi_b^0$ and $\Xi_b^-$ is that it decays {\it weakly} with a lifetime of the order of $1.4 \times 10^{-12}$ sec \cite{PDG} and therefore one can expect that the decays into the hyperon $\Xi$'s like $\Xi_b \to \Xi \gamma$, where $\Xi$ denotes the ground state of spin-parity ${1 \over 2}^+$, or $\Xi_b \to \Xi^* \gamma$, where $\Xi^*$ is a higher mass state, will have non-negligible branching ratios. The inclusive semileptonic $\Xi_b \to \ell \nu \Xi + X$ and non-leptonic $\Xi_b \to J/\Psi \Xi$ decays have been already observed \cite{PDG}.\par

One can consider decays of the type $\Lambda_b \to \Lambda (\to p \pi^-) \gamma$, e.g. $\Xi_b \to \Xi (\to \Lambda \pi) \gamma$. As for the case of the $\Lambda$, the final $\Xi$, being long-lived, will decay most of the time ouside the detector. Of course, one could also examine the decays to strongly decaying resonances, analogous to $\Lambda_b \to \Lambda^{*} (\to K^- p) \gamma$, as for example $\Xi_b \to \Xi^{*} (\to \Xi \pi) \gamma$.\par
 
In the present paper we want to underline that some radiative decays of the isospin doublet $\Xi_b^0$ and $\Xi_b^-$ will have nice signatures. The charged baryon $\Xi_b^-$ will decay into the mode $\Xi^- \gamma$, where ground state $\Xi^-\left({1 \over 2}^+\right)$, being charged, can be detected. 
On the other hand, the neutral mode $\Xi_b^0$ can decay into resonances $\Xi^{*0} \gamma$,  with $\Xi^{*0}$ decaying {\it strongly} into observable {\it charged} particles within the detector.\par

Of course, the number of b-flavored baryons produced at LHC will play an important role. A. Fridman and R. Kinnunen \cite{FK} have made an estimation of the number of b-baryons expected in $pp$ collisions at a c.m. energy of $\sim 16$ TeV and a luminosity $\mathcal{L} = 10^{32}\ \textrm{cm}^{-2}\ \textrm{s}^{-1}$. This can give an idea of the rough number of b-baryons that could be expected at LHC. They find the following number of b-baryons/year (summing over the different isospin states) :
\beq
\label{1e}
N(\Lambda_b) / N(\Sigma_b) / N(\Xi_b) \simeq (1.8 \pm 0.1) \times 10^{10} / (3.6 \pm 0.3) \times 10^{9} 
/ (2.0 \pm 0.2) \times 10^{9}
\eeq
Therefore, even if $\Lambda_b$ is expected to be produced more abundantly, the $\Xi_b$ baryons will also be produced in great numbers.\par

There is an important remark to be done at this point. In the heavy quark limit, there are two states made out of $bsq$ quarks (q = u, d), $|\Xi_{b3}>$ and $|\Xi_{b6}>$, corresponding to the two possible $SU(3)$ representations for the light cloud, that has flavor quantum numbers $sq$ (q = u, d), namely $\bf{\overline{3}}$ (antisymmetric in flavor, that contains $\Lambda_b$ and the doublet $\Xi_{b3}$) and $\bf{6}$ (symmetric in flavor, that contains the triplet $\Sigma_b$, the doublet $\Xi_{b6}$ and the singlet $\Omega_b$) (see for the case of $\Xi_c$, p. 1183 of \cite{PDG}). The physical states made out of $bsq$ (q = u, d) are linear combinations of the heavy quark limit states : 

\beq
|\Xi_b>\ = \cos \theta_b\ |\Xi_{b3}> + \sin \theta_b\ |\Xi_{b6}>
\label{1bise}
\eeq
\beq
\label{1bis2e}
|\Xi'_b>\ = -\sin \theta_b\ |\Xi_{b3}> + \cos \theta_b\ |\Xi_{b6}>
\eeq
where the angle $\theta_b$ is subleading in ${1 \over m_b}$, of the order of the degree (see for example the discussion in \cite{BLS} for the case of the couple $\Sigma_c$ and $\Sigma'_c$).\par

Because of Fermi-Dirac statistics, the wave function of $|\Xi_{b3}>$ and $|\Xi_{b6}>$ will be respectively antisymmetric and symmetric in the spin of the valence light quarks. In the quark model, the mass difference between both states is given by the Fermi hyperfine interaction, inversely proportional to the heavy quark mass. The mass difference is then expected to be of the order $M_{\Xi'_b}-M_{\Xi_b} \simeq {m_c \over m_b} (M_{\Sigma'_c}-M_{\Sigma_c}) \simeq 35\ MeV$.\par

Unlike the case of the pair $\Sigma_c, \Sigma'_c$, in the case of the b-flavored baryons, only the lowest state $\Xi_b$, that decays weakly, has been observed. The state $\Xi'_c$ has been seen decaying into $\Xi_c \gamma$, and the same is expected for the analogous b-flavored baryon $\Xi'_b$.\par

In the present paper we are concerned by the lowest, weakly decaying, state $\Xi_b$. From now on we follow very closely the scheme of the paper by F. Legger and T. Schietinger \cite{LS}, applying their formalism to the decays that are of interest to us in the present paper, $\Xi_b \to \Xi \gamma$ and $\Xi_b \to \Xi^* \gamma$.

In terms of operators at the leading order in $\alpha_s$ 

$$O_7 = {em_b \over 16\pi^2}\ \overline{s}\ \sigma_{\mu\nu}\ {1+\gamma_5 \over 2}\ b\ F^{\mu\nu}$$
\beq
\label{2e}
O'_7 = {em_b \over 16\pi^2}\ \overline{s}\ \sigma_{\mu\nu}\ {1-\gamma_5 \over 2}\ b\ F^{\mu\nu}
\eeq
the effective Hamiltonian for photon emission writes

\beq
\label{3e}
\mathcal{H}_{eff} = - 4\ {G_F \over \sqrt{2}}\ V_{ts}^* V_{tb}\left(C_7 O_7 + C'_7 O'_7\right)
\eeq

In the Standard Model, for zero light quark masses, the photon has a purely left-handed helicity, and the ratio 

\beq
\label{4e}
r = {C'_7 \over C_7}
\eeq
that measures the amount of right-handed over left-handed helicity photons, would vanish, $r = 0$. If one does not neglect the strange quark mass, in the Standard Model $r$ is of the order of the ratio of quark masses 

\beq
\label{5e}
r \simeq {m_s \over m_b}
\eeq
Of course, beyond the Standard Model, the Wilson coefficient $C'_7$ can have new contributions, and one can get other values for $r$, of great interest as a signal of possible NP. This is the main purpose of a measurement of the parameter $r$ through the photon polarization.\par

The photon asymmetry

\beq
\label{6e}
\alpha_\gamma = {{P(\gamma_L)-P(\gamma_R)} \over {P(\gamma_L)+P(\gamma_R)}}
\eeq
is given, at leading order in $\alpha_s$, in terms of (\ref{4e}), by the expression

\beq
\label{7e}
\alpha_\gamma = {{1-|r^2|} \over {1+|r^2|}}
\eeq

We use now the results of the angular momentum calculations of \cite{LS}, applied to the cases we are interested in.

\par \vskip 0.5 truecm

\section{The case \bf $\Xi_b \to \Xi \gamma$}

\par \vskip 0.5 truecm

For the decay into the ground state ${1 \over 2}^+\ \Xi_b \to \Xi \gamma$ we have the photon angular distribution (formula (10) of \cite{LS})

\beq
\label{8e}
{2 \over \Gamma} {d\Gamma \over d \cos \theta_\gamma} = 1 - \alpha_\gamma P_{\Xi_b} \cos \theta_\gamma
\eeq
where $P_{\Xi_b}$ is the $\Xi_b$ polarization and $\theta_\gamma$ is the angle between the $\Xi_b$-spin and the direction of the photon momentum as defined in the $\Xi_b$ rest frame.\par 

The hyperon $\Xi$ is long-lived, decays dominantly into the $\Lambda \pi$ mode, and has a clear signature in its charged mode $\Xi^- \to \Lambda (\to p\pi^-) \pi^-$. However, it is more difficult to detect than the $\Lambda \to p\pi^-$ case in $\Lambda_b$ radiative decay, since it involves two long-lived particles. We must underline that, even if the secondary decay $\Xi^- \to \Lambda \pi^-$ is not observed within the detector, the mode $\Xi_b^- \to \Xi^- \gamma$ is detectable, since the hyperon $\Xi^-$ is charged. 

\par \vskip 0.5 truecm

\section{The case \bf $\Xi_b \to \Xi^* \gamma$}

\par \vskip 0.5 truecm

Here $\Xi^*$ denotes resonances with flavor quantum numbers ssq (q = u, d). We will distinguish between the ground state $\XiÊ\left({3 \over 2}^+ \right)$, of mass $\sim$ 1530 MeV, and higher mass excited states.

\subsection{The decay into the ground state \bf $\Xi_b \to \Xi \left({3 \over 2}^+ \right) \gamma$}

In this case one could consider the decay chain

\beq
\label{9e}
\Xi_b^0 \to \Xi^{*0} (\to \Xi^-\pi^+) \gamma
\eeq
where $\Xi^{*}$ is the ground state $\Xi \left({3 \over 2}^+\right)$, that would decay strongly within the detector into the charged particles $\Xi^-\pi^+$.\par

However, the state $|\Xi_b>$ (\ref{1bise}) is very much dominated by the  state $|\Xi_{b3}>$, antisymmetric in flavor, since the mixing angle $\theta_b$ is expected to be small $\theta_b \sim {m_c \over m_b} \theta_c$, of the order of 1 degree \cite{BLS}.\par 

In the constituent quark model, the wave function of the decuplet state $|\Xi^{*}>$ is totally symmetric in flavor and in spin. Therefore, in the picture in which the decay $\Xi_b^0 \to \Xi^{*0} \gamma$ proceeds by the Penguin transition $b \to s \gamma$ and a $us$ pair of light quarks remain as spectators, the overlap between $|\Xi_b>$ and $|\Xi^{*}>$ will be very small due to the smallness of $\theta_b$. The overlap can be expected to be suppressed by a large factor due to this effect. Therefore, we expect a very small branching ratio for this mode, that probably will not be of much use for our purpose.\par
Notice that this argument does not follow for the decay into the ground state ${1 \over 2}^+\ \Xi_b \to \Xi \gamma$, because the final baryon wave function has a symmetric piece in flavor (and spin), but also another one that is antisymmetric.

\subsection{The decay to $\Xi$ resonances of higher mass}

One could also consider decays to higher mass resonances, like $\Xi_b \to \Xi \left({3 \over 2}^-\right) \gamma$. This resonance has a mass of 1823 MeV and decays strongly into $\Lambda \overline{K}$ or $\Sigma \overline{K}$. The mode $\Lambda \overline{K}$ is dominant, but has the drawback that the $\Lambda$ is long-lived and will mostly decay outside the detector. The decay $\Sigma \overline{K}$ has a smaller branching ratio, but could be observed in the charged mode $\Sigma^+ K^-$. This case is analogous to the clean proposal $\Lambda_b \to \Lambda^* \gamma \to pK^- \gamma$ of refs. \cite{LS}\cite{HKLS}, but cannot compete with it in number of events.

\section{Comment on the observables for the decay into $J = {3 \over 2}$ resonances}

The relevant angular distributions in the decays into $J = {3 \over 2}$ resonances $\Lambda_b \to \Lambda^*(\to K^-p) \gamma$ or $\Xi_b^0 \to \Xi^{*0}(\to \Xi^-\pi^+) \gamma$, $\Xi_b^0 \to \Xi^{*0}(\to \Sigma^+K^-) \gamma$ are given by eqs. (15) and (19) of \cite{LS}.
Taking as example the case $\Xi_b^0 \to \Xi^{*0}(\to \Xi^-\pi^+) \gamma$,

\beq
\label{10e}
{2 \over \Gamma} {d\Gamma \over d \cos \theta_\gamma} = 1 - \alpha_\gamma^{(3/2)} P_{\Xi_b} \cos \theta_\gamma
\eeq

\beq
\label{11e}
{2 \over \Gamma} {d\Gamma \over d \cos \theta_{\Xi}} = {1 \over 1 - {\alpha_{\Xi}^{(3/2)} \over 3}} \left[1 - \alpha_{\Xi}^{(3/2)} \cos^2 \theta_{\Xi} \right]
\eeq
where the superindex $3/2$ means that we are dealing with a resonance $\Xi^*$ of spin ${3 \over 2}$ and $\theta_{\Xi}$ denotes the polar angle between the $\Xi^{*0}$ flight direction and the $\Xi^-$ momentum in the $\Xi^{*0}$ rest frame.\par
 
As pointed out in \cite{LS}, the asymmetry parameters are given by eqs. (18) and (20) of \cite{LS} :

\beq
\label{12e}
\alpha_\gamma^{(3/2)} = {1-\eta \over 1+\eta}\ \alpha_\gamma
\eeq

\beq
\label{13e}
\alpha_{\Xi}^{(3/2)} = {\eta - 1 \over \eta + {1 \over 3}}
\eeq
where $\alpha_\gamma$ is the photon polarization (\ref{6e}) and $\eta$ is a parameter measuring the relative strength of the modulus squares of the $\lambda_{\Xi^{*}} = 3/2$ and $\lambda_{\Xi^{*}} = 1/2$ amplitudes $\mathcal{A}_{{\lambda_{\Xi^{*}}},{\lambda_\gamma}}$, eq. (17) of \cite{LS} in the decay $\Xi_b \to \Xi^{*0} \gamma$ :

\beq
\label{14e}
\eta = {|\mathcal{A}_{{\pm3/2},{\pm1}}|^2 \over |\textsl{A}_{{\pm1/2},{\pm1}}|^2}
\eeq

Combining (\ref{12e}) and (\ref{13e}) one gets \cite{LS} :

\beq
\label{15e}
\alpha_\gamma = {1 \over 2}  \alpha_\gamma^{(3/2)} \left[1 - {3 \over \alpha_{\Xi}^{(3/2)}}\right]
\eeq

Of course, eqs. (\ref{12e}), (\ref{13e}), (\ref{15e}) are useful if $\eta$, defined in (\ref{14e}) is different from $1$. Very interestingly, G. Hiller et al. \cite{HKLS} have demonstrated in LEET (Large Energy Effective Theory  \cite{CLOPR}) or, equivalently, in lowest order in SCET (Soft Collinear Effective Theory \cite{BFPS}), that the helicity-3/2 amplitude vanishes :

\beq
\label{16e}
\eta \simeq 0
\eeq

From (\ref{12e}), in the approximation (\ref{16e}), formula (\ref{10e}) becomes

\beq
\label{16bise}
{2 \over \Gamma} {d\Gamma \over d \cos \theta_\gamma} \simeq 1 - \alpha_\gamma P_{\Xi_b} \cos \theta_\gamma
\eeq

In all generality, independently of the value of $\eta$, it is illuminating to rewrite the observable $\alpha_\gamma^{(3/2)} P_{\Xi_b}$ in (\ref{10e}) in terms of the observable $\alpha_{\Xi}^{(3/2)}$. One obtains, 

\beq
\label{17e}
{2 \over \Gamma} {d\Gamma \over d \cos \theta_\gamma} =  1 + {2 \over 3} {\alpha_{\Xi}^{(3/2)} \over {1 - {\alpha_{\Xi}^{(3/2)} \over 3}}} \alpha_\gamma P_{\Xi_b} \cos \theta_\gamma
\eeq

We observe that in both angular distributions for the cases $\Xi$ (\ref{8e}) and $\Xi^*$ (\ref{16bise}) or (\ref{17e}) appears the product

\beq
\label{18e}
\alpha_\gamma P_{\Xi_b}
\eeq
Therefore, to determine $\alpha_\gamma$ in either case we need experimental information on the polarization $P_{\Xi_b}$. On the other hand, the parameter $\alpha_{\Xi}^{(3/2)}$ in (\ref{17e}) is an independent observable that can be measured by the angular distribution (\ref{11e}). Of course, this statement applies also to the corresponding $\Lambda_b$ decays. \par 

\section{Comments on $\Lambda_b$ and $\Xi_b$ polarization}

Let us now concentrate on the question of the theoretical estimation or the possibility of measurement of the transverse polarization $P_{\Xi_b}$. The most studied b-baryon in the literature has been $\Lambda_b$. We will make some comments on the $\Xi_b$ case below.\par 

\subsection{Summary and comments on the $\Lambda_b$ case}

Let us first summarize the different results or hints in the most studied $\Lambda_b$ case, based on the papers that we have quoted above. We repeat here the different attemps or guesses indicated in the literature, adding also some personal remarks that we consider important.

\par \vskip 0.2 truecm

(i) Following general arguments in hadronic collisions \cite{L}, $P_{\Lambda_b}$ could be greater than $20 \%$. For $P_{\Xi_b}$ one could guess something of the same order, although the statistics is here smaller, as we see from (\ref{1e}), since a strange quark has to be created.

\par \vskip 0.2 truecm

(ii) From perturbative QCD \cite{DG} one gets a quark transversal polarization 

\beq
\label{19e}
P_b \propto \alpha_s m_b
\eeq 
The b-quark having a rather large mass, one can expect a polarization of $\mathcal{O}(10 \%)$, a large fraction of which is transferred to the $\Lambda_b$ polarization \cite{FP}. Intuitively this can be easily understood, since in the heavy quark limit one has $P_{\Lambda_b} = P_b$ \cite{MS}.

\par \vskip 0.2 truecm

(iii) As studied in detail in \cite{HLS}\cite{ACL} one could {\it measure} $P_{\Lambda_b}$ through the clean decay chain $\Lambda_b \to \Lambda (\to p\pi^-) J/\Psi (\to \ell^+\ell^-)$. Although the primary decay $\Lambda_b \to \Lambda J/\Psi$ violates parity, the $\Lambda$ decay also violates parity, with known asymmetry parameter $\alpha$. This provides a sufficient number of observable correlations to determine $P_{\Lambda_b}$. The duplication of unknowns from parity violation in the primary decay is compensated by the parity violation of the $\Lambda$ decay product.\par 
We encounter here the same problem underlined above that the $\Lambda$ will decay mostly outside the detector. This measurement would be difficult.\par 

We must emphasize however that the branching ratio has already been measured at the Tevatron \cite{PDG} :

\beq
\label{19bise}
\textrm{BR}^{exp}(\Lambda_b \to \Lambda J/\Psi) = (4.7 \pm 2.8) \times 10^{-6}
\eeq

\par \vskip 0.2 truecm

(iv) On the other hand, integrating over the azimuthal angle, one has the differential decay rate for $\Lambda_b \to \Lambda J/\Psi$ \cite{ACL}

\beq
\label{20e}
{2 \over \Gamma} {d \Gamma \over d \cos \theta} = 1 + \alpha_{\Lambda_b} P_{\Lambda_b} \cos \theta
\eeq
where $\theta$ is the zenithal angle of the hyperon $\Lambda$ in the $\Lambda_b$ rest frame, and $\alpha_{\Lambda_b}$ is the {\it parity-violating} helicity asymmetry of the decay $\Lambda_b \to \Lambda J/\Psi$ 

\beq
\label{21e}
\alpha_{\Lambda_b} = {|\mathcal{M}_{\Lambda_b}(+ {1 \over 2})|^2-|\mathcal{M}_{\Lambda_b}(- {1 \over 2})|^2 \over |\mathcal{M}_{\Lambda_b}(+ {1 \over 2})|^2+|\mathcal{M}_{\Lambda_b}(- {1 \over 2})|^2}
\eeq

Theoretical calculations of this asymmetry within the factorization hypothesis have been performed \cite{ACL}. These calculations make use of the form factors appearing in the amplitude 

\beq
\label{22e}
<\Lambda|\bar{s}\gamma_\mu(1-\gamma_5)b|\Lambda_b> = \bar{u}_\Lambda\left\{\left[F_1(q^2)+{/\hskip - 2 truemm v}F_2(q^2)\right]\gamma_\mu(1-\gamma_5)\right\}u_{\Lambda_b}
\eeq
$\left(v = {P_{\Lambda_b} \over M_{\Lambda_b}}\right)$ estimated using QCD Sum Rules and HQET \cite{HY}. This leads, within factorization, to the theoretical estimate for the asymmetry parameter $\alpha_{\Lambda_b}$ (\ref{21e}),

\beq
\label{23e}
\alpha_{\Lambda_b} \simeq 80 \%
\eeq
pointing to a realistic possibility for the measurement of the $\cos \theta$ term in (\ref{20e}), and therefore to a possible estimate of $P_{\Lambda_b}$.\par  

In \cite{LAC} it is underlined that the estimation of the branching ratio using factorization $\textrm{BR}^{fact}(\Lambda_b \to \Lambda J/\Psi) = 3.1 \times 10^{-6}$ is in agreement, within errors, with the present experimental result (\ref{19bise}). This is encouraging, and since in the decay $\Lambda_b \to \Lambda J/\Psi$ one is dealing with a heavy hadron $\Lambda_b$ as initial state, the production of a light hadron $\Lambda$ and the emission of a meson $J/\Psi$ made out of a pair of quarks of identical mass, the estimation of its amplitude by factorization \cite{ACL}\cite{LAC} could be considered the lowest order within QCD Factorization \cite{BBNS}. Although difficult, one could in principle undertake the task of improving the factorization result for the amplitudes $\Lambda_b \to \Lambda J/\Psi$, and therefore for the polarization (\ref{23e}), and control its hadronic uncertainties by using a theoretical scheme such as QCD Factorization. 

\par \vskip 0.2 truecm

(v) In order to try to have information on the $\Lambda_b$ polarization, one could consider the non-leptonic decay $\Lambda_b \to \Lambda^{*} J/\Psi$, parallel to the promising Penguin mode \cite{LS}\cite{HKLS} $\Lambda_b \to \Lambda^{*} \gamma$, where $\Lambda^{*}$ denotes an excited state decaying strongly within the detector into the clean mode $p K^-$. The primary decay $\Lambda_b \to \Lambda^{*} J/\Psi$ violates parity, but the decay of the $\Lambda^{*}$ conserves parity. Therefore, one is not here in the situation of determining experimentally, by angular correlations, the corresponding parity-violation asymmetry parameter (or parameters) $\alpha_{\Lambda_b}^{*}$ (different in the decay $\Lambda_b \to \Lambda^{*} J/\Psi$ considered here than in the case $\Lambda_b \to \Lambda J/\Psi$), and determine in a model-independent way the crucial $\Lambda_b$ polarization, the main unknown of the problem. One is forced also here to introduce some theory-dependent input on the primary decay in order to have an estimate of the parity-violating asymmetry and therefore of the $\Lambda_b$ polarization.\par 

However, even if we do not expect to have enough information to determine the $\Lambda_b$ polarization in this case, a further study of the angular correlations in the decay chain $\Lambda_b \to \Lambda^{*} (\to pK^-) J/\Psi (\to \ell^+\ell^-)$ (where $\Lambda^{*}$ is an excited state of spin $3 \over 2$) could be worth to be undertaken.

\par \vskip 0.5 truecm

\subsection{Comments on $\Xi_b$}

Let us now make some comments on the polarization of $\Xi_b$.\par
In \cite{HLS}, the possibility of measuring the $\Xi_b^0$ polarization is proposed through the decay chain $\Xi_b^0 \to \Lambda (\to p\pi^-) J/\Psi (\to \ell^+\ell^-)$. However, as pointed out already in \cite{HLS} the decay $\Xi_b^0 \to \Lambda J/\Psi$ proceeds through the Cabibbo suppressed quark decay $b \to dc\overline{c}$ as compared to the Cabibbo-allowed decay $b \to sc\overline{c}$ that governs $\Lambda_b \to \Lambda J/\Psi$. Notice that the Cabibbo-suppressed decay $\Xi_b^0 \to \Lambda J/\Psi$ is different from the ones $\Xi_b^0 \to \Xi J/\Psi$ or $\Xi_b^0 \to \Xi^{*} J/\Psi$ on which we are interested here.\par 

Following what we have learned from the $\Lambda_b$ in the preceding Subsection concerning points (i)-(v), the same comments apply to $\Xi_b^0 \to \Xi J/\Psi$ or $\Xi_b^0 \to \Xi^{*} J/\Psi$.\par

As pointed out above, the equivalent decay to (iii) to measure the $\Xi_b$ polarization would be $\Xi_b \to \Xi (\to \Lambda\pi) J/\Psi (\to \ell^+\ell^-)$ that, for the charged case $\Xi_b^{-}$ it will have a very clean signature $\Xi^- \to \Lambda (\to p\pi^-)\pi^-$, although hardly observable within the detector.\par

The comments (iv) and (v) apply respectively to the decays $\Xi_b \to \Xi J/\Psi$
and $\Xi_b \to \Xi^* J/\Psi$. There is already preliminary data on the first mode, and it could be very interesting to have experimental information on both. We must underline that the comment (v) applies to the example of the decay chain $\Xi_b^0 \to \Xi^{*0}\left({3 \over 2}^-\right) (\to \Sigma^+K^-) J/\Psi (\to \ell^+\ell^-)$, because the secondary decays conserve parity.

\section{Conclusion}

To conclude, in the present paper we have investigated the possibility of having information on the photon polarization of the Penguin transition $b \to s \gamma$ through the weak decays of the b-flavored baryon $\Xi_b$. We have made extensive use of work done in the literature for the case of the $\Lambda_b$. We consider the decays into ground state baryons of the octet $\Xi_b^- \to \Xi^-\left({1 \over 2}^+\right) \gamma$ and into $\Xi_b^0 \to \Xi^{*0} \gamma$, where the resonances $\Xi^{*0}$ can decay strongly into charged particles within the dectector. We emphasize the importance of having information on the initial $\Xi_b$ (or $\Lambda_b$) polarization, and examine, case by case, which kind of theoretical input is necessary to arrive to the main goal, the estimation of the photon polarization.

\par \vskip 1.0 truecm

\noindent {\large \bf Acknowledgements}

\par \vskip 0.5 truecm

This work has been supported in part by the EU Contract No. MRTN-CT-2006-035482, FLAVIAnet. We are indebted to Alain Le Yaouanc for dicussions and information on the literature on $\Lambda_b$ radiative decay.

\par 


\vskip 0.5 truecm

\begin{thebibliography}{99}

\bibitem{AGS} D. Atwood, M. Gronau and A. Soni, hep-ph/9704272, Phys. Rev. Lett. \textbf{79}, 185 (1997).

\bibitem{GP} M. Gronau, Y. Grossman, D. Pirjol and A. Ryd, Phys. Rev. Lett. \textbf{88}, 051802 (2002); M. Gronau and D. Pirjol, Phys. Rev. D \textbf{66}, 054008 (2002).

\bibitem{GKS} M. Gremm, F. Kr$\ddot {\textrm{u}}$ger and L.M. Sehgal, Phys. Lett. B \textbf{355}, 579 (1995).

\bibitem{MR} T. Mannel and S. Recksiegel, Acta Phys. Pol. B \textbf{28}, 2489 (1997); T. Mannel and S. Recksiegel, J. Phys. G : Nucl. Part. Phys. \textbf{24}, 979 (1998).

\bibitem{HK} G. Hiller and A. Kagan, Phys. Rev. D \textbf{65}, 074038 (2002). 

\bibitem{LS} F. Legger and T. Schietinger, hep-ph/0605245, Phys. Lett. B \textbf{645}, 204 (2007), Erratum-ibid. B \textbf{647}, 527 (2007).

\bibitem{HKLS} G. Hiller, M. Knecht, F. Legger and T. Schietinger, hep-ph/0702191, Phys. Lett. B \textbf{649}, 152 (2007). 

\bibitem{PDG} Review of Particle Physics (Particle Data Group), Phys. Lett. B \textbf{667}, 1 (2008).

\bibitem{FK} A. Fridman and R. Kinnunen, CERN-PPE/93-61, Contribution to the 5th International Symposium on Heavy Flavor Physics, Montr\'eal, July 1993.

 \bibitem{BLS} G. Glenn Boyd, Ming Lu and M.J. Savage, hep-ph/961244, Phys. Rev. \textbf{D55}, 5474 (1997).

\bibitem{CLOPR} J. Charles, A. Le Yaouanc, L. Oliver, O. P\`ene and J.-C. Raynal, Phys Rev. D \textbf{60}, 014001 (1999).

\bibitem{BFPS} C.W. Bauer, S. Fleming, D. Pirjol and I.W. Stewart, Phys Rev. D \textbf{63}, 114020 (2001); M. Beneke, A.P. Chapovsky, M. Diehl and T. Feldmann, Nucl. Phys. B \textbf{643}, 431 (2002).

\bibitem{L} E. Leader, {\it Spin in Particle Physics}, Cambridge University Press.

\bibitem{DG} W.G.D. Dharmaratna and G.R. Goldstein, Phys. Rev. D \textbf{53}, 1073 (1996).

\bibitem{FP} A.F. Falk and M.E. Peskin, Phys. Rev. D \textbf{49}, 3320 (1996).

\bibitem{MS} T. Mannel and G.A. Schuler, Phys. Lett. B \textbf{279}, 194 (1992).

\bibitem{HLS} J. H\v{r}ivn\'{a}\v{c}, R. Lednick\'{y} and M. Smi\v{z}ansk\'{a}, hep-ph/9405231, J. Phys. G : Nucl. Part. Phys. \textbf{21}, 629 (1995).

\bibitem{ACL} Z.J. Ajaltouni, E. Conte and O. Leitner, hep-ph/0412116, Phys. Lett. B \textbf{614}, 165 (2005); hep-ph/0412131, presented at 10th International Conference on Structure of Baryons (Baryons 2004), Palaiseau, France, October 2004, Nucl. Phys. A \textbf{755}, 435 (2005). 

\bibitem {HY} C.S. Huang and H.G. Yan, Phys. Rev. D \textbf{59}, 114022 (1999) [Erratum-ibid. D \textbf{61}, 039901 (2000).

\bibitem {LAC} O. Leitner, Z.J. Ajaltouni and E. Conte, hep-ph/0602043. 

\bibitem {BBNS} M. Beneke, G. Buchalla, M. Neubert and C.T. Sachrajda, hep-ph/0006124, Nucl. Phys. B \textbf{591}, 313 (2000). 

\end{thebibliography}
\end{document}